\documentstyle[12pt]{article} 
\newcommand{\be}{\begin{equation}}
\newcommand{\ee}{\end{equation}}
\newcommand{\ba}{\begin{array}}
\newcommand{\ea}{\end{array}}
\newcommand{\bc}{\begin{center}}
\newcommand{\ec}{\end{center}}

\newcommand{\apb}[1]{Ann. of Phys. (N.Y.) {\bf #1} }
\newcommand{\npb}[1]{Nucl. Phys. {\bf #1}}
\renewcommand{\theequation}{\thesection.\arabic{equation}}

\begin{document}

\title{
Variational approximation for two-time correlation functions in
$\Phi^4$ theory :   \\
optimization of the dynamics
}
\author{
C\'ecile Martin
\\
\\
{\normalsize
Division de Physique Th\'eorique\thanks{Unit\'e de Recherche des Universit\'es
Paris XI et Paris VI associ\'ee au C.N.R.S},}\\
{\normalsize  Institut de Physique Nucl\'eaire,} \\
{\normalsize F-91406 , Orsay Cedex, France }
}

\date{}

\maketitle
\vspace*{1cm}
\begin{abstract}

 We apply the time-dependent variational principle of Balian and
V\'en\'eroni  to the $ \Phi^4$ theory.
An appropriate  parametrization for the variational objects  allows us
to write coupled dynamical equations from which we derive
approximations for  the two-time correlation
functions involving two, three or four field operators.
\end{abstract}


\noindent IPNO/TH 95-22
\newpage

\section{Introduction}

\setcounter{equation}{0}

 A basic tool of quantum field theory is the generating functional for
multi-time correlation functions. In this paper, we show how the
time-dependent variational principle of Balian and Veneroni \cite{1} can be
implemented in the context of quantum field theory out of equilibrium.
{}From this approach, we derive approximate non perturbative
dynamical equations for expectation values and
correlation functions of local and composite operators. These evolution
equations could be useful to evaluate the fluctuations of the matter
density in cosmological scenarios \cite{2}.
They could  also provide informations
about the non-equilibrium dynamics during the expansion of the
quark-gluon plasma \cite{3,3a}. We consider here the case of a self-interacting
scalar field in a Minkowski metric described by the  hamiltonian
  $
H= \int d^d x \: {\cal H}(\vec x)$ with
\be \label{1.1}
{\cal H} (\vec x) = \frac{1}{2} \: \Pi^2(\vec x) + \frac{1}{2} \:
 (\vec \nabla \Phi
(\vec x))^2 + \frac{m_0^2}{2} \: \Phi^2(\vec x) + \frac{b}{24}
\: \Phi^4(\vec x) \ . \ee
We work in $d$ spatial dimensions. The constants $m_0$ and $b$ are
respectively the bare mass and the bare coupling constant.

\section{Variational evaluation of the generating functional  }

\setcounter{equation}{0}

Let us define an operator $A(J,t)$, functional of the sources $J$
and function of the time $t$, having the form :
\be \ba{lll}\label{2.1}
A(J,t) & {\displaystyle = T \exp i \int_t^{\infty} dt' \: \left\{
\int d^dx \: \: J^{\Phi}(\vec x,t') \Phi^H(\vec x,t',t) +
J^{\Pi}(\vec x,t') \Pi^H(\vec x,t',t) \right. } \\
& {\displaystyle  \left.
+ \int d^dx \: d^dy \: \: J^{\Phi \Pi}(\vec x,\vec y,t') \:
\left( \Phi^H(\vec x,t',t) \Pi^H(\vec y,t',t) + \Pi^H(\vec y,t',t)
\Phi^H(\vec x,t',t) \right) \right. } \\
& {\displaystyle \left.
+J^{\Phi \Phi}(\vec x, \vec y, t') \:
\Phi^H(\vec x, t',t) \Phi^H(\vec y,t',t)
+J^{\Pi \Pi}(\vec x, \vec y,t') \: \Pi^H(\vec x,t',t) \Pi^H(\vec y,t',t)
\right\} }
\ea  \ , \ee
where $T$ is the time-ordering operator and
 $\Phi^H(\vec x,t',t)$ and $\Pi^H(\vec x,t',t)$
are respectively the field and momentum operators defined in the Heisenberg
representation. They satisfy the boundary conditions :
$\Phi^H(\vec x,t,t)=\Phi(\vec x)$ and $\Pi^H(\vec x,t,t)=\Pi(\vec x)$,
$\Phi(\vec x)$ and $\Pi(\vec x)$ being the operators in the Schr\"odinger
representation. The local as well as bilocal sources $J(t')$
are turned on between $t'=t$ and
$ t'=+ \infty$.
The generating functional for the causal Green functions writes
\be \label{2.2}
Z(J,t_0)  = Tr \left(D(t_0) \: A(J,t_0) \right) \ , \ee
$D(t_0)$ being the initial state.
The statistical  operator $D(t_0)$ may represent a thermal equilibrium or a
non-equilibrium situation. At zero temperature, it reduces to a projection
operator. Besides the sources $J$, the generating functional $Z$ depends on
the initial time $t_0$.
We  want  to evaluate the functional derivatives of
\be \label{2.3}
W(J,t_0) = -i \: \ln Z(J,t_0) \ , \ee
since $W(J,t_0)$ is the generating functional for the
connected Green functions.
Its expansion in powers of the sources writes :
\be \ba{lllll} \label{2.3a}
& {\displaystyle W(J
, t_0)= -i \:  n_0 +
 \int_{t_0}^{+ \infty} dt' \: \{
\int d^dx \: J_{\Phi}(\vec x,t') \: \varphi(\vec x,t') +
\int d^dx \: d^dy \: J_{\Phi \Phi}(\vec x,\vec y,t') \: G(\vec x, \vec y,t') +
... \}
} \\
& {\displaystyle + \frac{i}{2} \: \int \int_{t_0}^{+ \infty} dt' \: dt'' \:
\{
\int d^dx_1 \: d^dx_2 \: J_{\Phi}(\vec x_1,t') \: J_{\Phi}(\vec x_2,t'') \:
C^2_{\Phi \Phi}(\vec x_1,\vec x_2,t',t'') + ...  \}  } \\
& {\displaystyle + ... } \ea \ ,  \ee
where  $n_0=Tr D(t_0)$ ( we choose not
to normalize $D(t_0)$).
We  use the following notations for the expectation values of
one and two field operators :
\be \label{2.3b}
\varphi(\vec x,t) = \frac{1}{n_0} \: Tr \left( \Phi^{H}(\vec x,t,t_0)
\: D(t_0) \right) \ , \ee
\be \label{2.3c}
G(\vec x, \vec y,t)=\frac{1}{n_0} \: Tr \left( \Phi^{H}(\vec x,t,t_0)
\: \Phi^{H}(\vec y,t,t_0) \: D(t_0) \right) - \varphi(\vec x,t) \:
\varphi(\vec y,t) \ . \ee
The two-time causal functions between two field operators are
defined according to :
\be \label{2.3d}
C^2_{\Phi \Phi}(\vec x_1,\vec x_2,t_1,t_2)= \frac{1}{n_0} \: Tr \left(
T \: \Phi^{H}(\vec x_1,t_1,t_0) \: \Phi^{H}(\vec x_2,t_2,t_0) \: D(t_0) \right)
- \varphi(\vec x_1,t_1) \: \varphi(\vec x_2,t_2) \ . \ee
Similarly, we define $C_{\Phi \Pi}^2$, $C_{\Pi \Pi}^2$.
We define also the three-point and four-point two-time correlation
function $ C^3$ and $C^4$ according to :
\be \label{2.3e} \ba{ll}
\frac{1}{i} & {\displaystyle
 \frac{\delta^2 W}{\delta J_{\Phi}(\vec x,t')
\delta J_{\Phi \Phi}(\vec y, \vec z,t")} \big\vert_{J=0} \equiv
C^3(\vec x, \vec y, \vec z, t', t")
} \\
& {\displaystyle
+ \varphi(\vec y,t") \: \left( C^2(\vec x,\vec z,t',t") + \varphi(\vec x,t')
\: \varphi(\vec z, t") \right)
+ \varphi(\vec z,t") \: \left( C^2(\vec x,\vec y,t',t") + \varphi(\vec x,t')
\: \varphi(\vec y, t") \right)
} \ , \ea \ee
\be \label{2.3f} \ba{lll}
\frac{1}{i} & {\displaystyle
\frac{\delta^2 W}{\delta J_{\Phi \Phi}(\vec x,\vec y,t')
\delta J_{\Phi \Phi}(\vec z, \vec u,t")} \big\vert_{J=0} \equiv
C^4(\vec x, \vec y, \vec z, \vec u,t', t")
} \\
& {\displaystyle
+ \left( C^2(\vec x, \vec z, t',t") + \varphi(\vec x,t') \:
\varphi(\vec z,t") \right) \:
\left( C^2(\vec y, \vec u, t',t") + \varphi(\vec y,t') \:
\varphi(\vec u,t") \right)
} \\
& {\displaystyle
+ \left( C^2(\vec x, \vec u, t',t") + \varphi(\vec x,t') \:
\varphi(\vec u,t") \right) \:
\left( C^2(\vec y, \vec z, t',t") + \varphi(\vec y,t') \:
\varphi(\vec z,t") \right)
} \ .  \ea \ee

We will use the time-dependent variational principle of Balian and
Veneroni \cite{1} to obtain  an approximation for our quantity of interest,
 $Tr(D(t_0)A(J,t_0))$.
We  define therefore the action   functional
\be \label{2.4}
{\cal Z} \left({\cal A}(t), {\cal D}(t) \right) =
Tr \left({\cal A}(t_0) \: {\cal D}(t_0) \right) + {\cal Z}_{dyn} \ , \ee
with
\be \label{2.5}
{\cal Z}_{dyn} = Tr \: \int_{t_0}^{\infty}  dt \: \: {\cal D}(t) \:
\left\{ \frac{d {\cal A}(t)}{dt}
- i \left[{\cal A}(t), \int d^dx {\cal H}(\vec x)
\right] + i {\cal A}(t) \: \left( \sum_j J_j(t) Q_j \right)  \right\} \ , \ee
 where we have written in a compact form
the term which involves the sources $J$ :
\be \ba{lll} \label{2.6}
\sum_j J_j(t) Q_j =& {\displaystyle
 \int d^dx \: \left(J^{\Phi}(\vec x,t) \Phi(\vec x) +
J^{\Pi}(\vec x,t) \Pi(\vec x) \right) } \\
& {\displaystyle + \int d^dx \: d^dy  \: J^{\Phi \Pi}(\vec x,\vec y,t)
\left( \Phi(\vec x)\Pi(\vec y) + \Pi(\vec y) \Phi(\vec x)  \right) } \\
& {\displaystyle  + \int d^dx \: d^dy \:
\left(  J^{\Phi \Phi}(\vec x,\vec y,t) \Phi(\vec x)
\Phi(\vec y) + J^{\Pi \Pi}(\vec x,\vec y,t) \Pi(\vec x) \Pi(\vec y) \right) }
\ . \ea \ee
The variational quantities of the  functional ${\cal Z}$, which  depends on
the sources $J$, are the    observable-like and density-like
operators   ${\cal A}(t)$ and ${\cal D}(t)$.
We look for the stationarity of
 ${\cal Z}$ with respect to variations of ${\cal A}(t)$ and  ${\cal D}(t)$
subject to the boundary conditions :
 ${\cal A}(t=+ \infty) =  1$ and
${\cal D} (t_0) = D(t_0)$, where $D(t_0)$ is the initial statistical operator
 which
we assume to be given and equal to a Gaussian density matrix.
 The generating functional for the connected Green functions will
be approximated by $W(J,t_0)=-i \: \ln {\cal Z}_{st}$.

We restrict ourselves to   trial operators ${\cal A}(t)$ and  ${\cal D}(t)$
which are exponentials of quadratic
and linear forms of the field operators $\Phi(\vec x)$ and $\Pi(\vec x)$
(which we shall
loosely call Gaussian operators).
 Let us introduce the operators
${\cal T}_b = {\cal A}(t) \: {\cal D} (t)$ and
${\cal T}_c={\cal D}(t) \: {\cal A} (t)$ \cite{4}.
We define the following expectation values :
\be \label{2.7}
n(t)=Tr {\cal T}_b=Tr{\cal T}_c \ , \ee
\be \label{2.8}
\varphi_b(\vec x,t)=\frac{1}{n(t)} \: Tr({\cal T}_b \Phi(\vec x)) \ , \ee
\be \label{2.9}
\pi_b(\vec x,t) = \frac{1}{n(t)} \: Tr({\cal T}_b \Pi(\vec x)) \ , \ee
\be \label{2.10}
G_b(\vec x,\vec y,t)= \frac{1}{n(t)} \: Tr({\cal T}_b \tilde \Phi(\vec x)
\tilde \Phi(\vec y)) \ , \ee
\be \label{2.11}
S_b(\vec x,\vec y,t) = \frac{1}{n(t)} \: Tr({\cal T}_b \tilde \Pi(\vec x)
\tilde \Pi(\vec y) ) \ , \ee
\be \label{2.12}
T_b(\vec x, \vec y,t) = \frac{1}{n(t)} \: Tr({\cal T}_b
( \tilde \Phi(\vec x) \tilde \Pi(\vec y) +
\tilde \Pi(\vec y) \tilde \Phi(\vec x) ))
\ , \ee
with
$\tilde \Phi(\vec x) = \Phi(\vec x)-\varphi_b(\vec x,t)$ and
$\tilde \Pi(\vec x) =\Pi(\vec x) -\pi_b(\vec x,t)$.
Similarly we define the expectation values associated with
${\cal D}(t)  {\cal A}(t)$ and indexed by  $c$ by replacing
 ${\cal T}_b$ by
${\cal T}_c$ in (\ref{2.8})-(\ref{2.12}).
For  ${\cal D}(t)=1$,
one has $\varphi_{b}(\vec x,t) =\varphi_{c}(\vec x,t)=\varphi_{a}(\vec x,t) $
(and similar definitions for $\pi_a, G_a, S_a, T_a$).
For ${\cal A}(t) =1$, one has
$\varphi_{b}(\vec x,t) =\varphi_{c}(\vec x,t)=\varphi_{d}(\vec x,t) $
 (and similar definitions for  $\pi_d, G_d, S_d, T_d$).
A Gaussian operator is completely characterized by a set
$\{ \varphi, \pi, G, S, T \}$. It will be  convenient to introduce a
 vector $\alpha(\vec x,t)$
\be \label{2.13}
\alpha(\vec x, t) = \pmatrix{ \varphi(\vec x,t) \cr
-i \: \pi(\vec x,t) \cr} \ , \ee
and a matrix  $\Xi (\vec x,
\vec y, t)$
\be \label{2.14}
\Xi (\vec x, \vec y,t) = \pmatrix{
2 \: G(\vec x, \vec y,t) & -i \: T(\vec x, \vec y,t) \cr
-i \: T(\vec y,\vec x,t) & -2 \: S(\vec x, \vec y,t) \cr } \ . \ee
We give in appendix the expressions of
$\alpha_b, \alpha_c, \Xi_b$ and $ \Xi_c$ as functions of the independent
variational quantities
$\alpha_a,  \Xi_a$ and $\alpha_d,  \Xi_d$ which characterize ${\cal A}$ and
${\cal D}$ respectively.

With this choice of the trial  spaces for ${\cal A}(t)$ and
${\cal D}(t)$ , the Wick theorem allows us to express
the functional  ${\cal Z}$ in the form :
\be \ba{ll} \label{2.15}
 {\cal Z}=
& {\displaystyle n(t_0) + \int_{t_0}^{+ \infty} dt \: \left[
\frac{dn}{dt} \vert_{{\cal D}(t)=cte} - i \: n(t) \: \int d^dx \:
\left( {\cal E}_c(\vec x,t) - {\cal E}_b(\vec x,t) \right) \right. } \\
& {\displaystyle \left. + i \: n(t) \: \int d^dx_{1} d^dx_2 \:
{\cal K}_c (\vec x_1, \vec x_2,t) \right] } \ea \ , \ee
where the energy density ${\cal E}(\vec x, t)=Tr \left( {\cal D}(t) \:
{\cal H}(\vec x, t) \right)$ is given, for the Hamiltonian density
(\ref{1.1}), by :
\be \ba{lll} \label{2.16}
{\cal E}(\vec x,t)=
& {\displaystyle \frac{1}{2} \: \pi^2(\vec x,t) + \frac{1}{2} \:
(\vec \nabla \varphi )^2 + \frac{m_0^2}{2} \: \varphi^2(\vec x,t) +
\frac{b}{24} \: \varphi^4(\vec x,t) } \\
& {\displaystyle + \frac{1}{2} \: S(\vec x, \vec x,t) - \frac{1}{2} \:
\Delta G(\vec x, \vec y,t) \vert_{x=y} + \frac{m_0^2}{2} \:
G(\vec x, \vec x,t) } \\
& {\displaystyle + \frac{b}{8} \: G^2(\vec x, \vec x,t) +
\frac{b}{4} \: \varphi^2(\vec x,t) \: G(\vec x, \vec x,t) } \ . \ea \ee
The last term of  the functional (\ref{2.15}),
$K_c = Tr \left( {\cal T}_c \: \sum_{j} J_j(t) \: Q_j \right) =
\int d^dx_1 d^dx_2 \: {\cal K}_c(x_1,x_2,t)$, involves the sources $J$ and
the expectation values indexed by $c$.

\section{Dynamical equations}

\setcounter{equation}{0}

By varying the expression (\ref{2.15}) with respect to $n_d, \alpha_d, \Xi_d,$
with the boundary conditions
\be \label{3.1}
n_d(t_0)=n_0 \: \: ,
\: \: \alpha_d(t_0)=\alpha_0 \: \: , \: \: \Xi_d(t_0)=\Xi_0 \ , \ee
where
$n_0, \alpha_0$ et $\Xi_0$ characterize  the initial Gaussian state $D(t_0)$,
we obtain the  evolution equations for $n_a, \alpha_a$ and
$ \Xi_a$. Integrating
(\ref{2.15}) by parts and varying with respect to $n_a, \alpha_a, \Xi_a$
with the boundary conditions
\be \label{3.2}
n_a(t=+ \infty)=1 \: \: ,
\: \: \alpha_a(t=+ \infty)=0 \: \: ,\: \: \Xi_a^{-1}(t=+ \infty)=0
\ , \ee
we obtain the evolution equations for $n_d, \alpha_d$ and $ \Xi_d$.
In general the evolution equations for $n_d, \alpha_d, \Xi_d$ and
those for $n_a,
\alpha_a,   \Xi_a$ are coupled.
The solutions  $n_d, \alpha_d, \Xi_d$ and $n_a, \alpha_a, \Xi_a$  depend
on the sources.

By combining the evolution equations for $n_d, \alpha_d, \Xi_d$
with those for  $n_a, \alpha_a, \Xi_a$,
the dynamical equations for the expectation values
$\alpha_b, \Xi_b$ and $ \alpha_c, \Xi_c$ can then be written in the following
 compact form :
\be \label{3.3}
i \: \dot \alpha_b = \tau \: w^b \ , \ee
\be \label{3.4}
i \: \dot \alpha_c = \tau \: \left( w^c - w^c_K \right) \ , \ee
\be \label{3.5}
i \: \dot \Xi_b = 2 \left( \Xi_b \: {\cal H}^b \: \tau - \tau \:
{\cal H}^b \: \Xi_b \right) \ , \ee
\be \label{3.6}
i \: \dot \Xi_c = 2 \: \left( \Xi_c \: \left( {\cal H}^c -{\cal I}^c_K
\right) \: \tau - \tau \: \left( {\cal H}^c - {\cal I}^c_K \right) \: \Xi_c
\right) \ , \ee
where $\tau$ is the $2 \times 2 $ matrix
\be \label{3.7}
\tau= \pmatrix{ 0 & 1 \cr -1 & 0 \cr } \ . \ee
  The  vector $w^b$  and the  matrix
${\cal H}^b$ are defined from the variation  of
 $<H>_b=Tr({\cal T}_b H)$  :
\be \label{4.8}
\delta <H>_b = \int d^dx \: \tilde w^b(\vec x,t) \: \delta \beta_b(\vec x,t)
-\frac{1}{2} \: \int d^dx \: d^dy \: tr {\cal H}^b(\vec x, \vec y,t) \:
\delta \Xi_b(\vec x, \vec y,t) \ .  \ee
Similarly,  $w^c, {\cal H}^c$ and
$w^c_K, {\cal I}^c_K$ are defined from the variation of
$<H>_c=Tr({\cal T}_c H)$ and from the variation of the  source term $K_c$.

Therefore, using a convenient parametrization for the two variational
objects ${\cal D}(t)$ and ${\cal A}(t)$, we have been able to obtain
dynamical equations  in a very compact form   even
for finite values of the sources.
The dynamical equations for $\alpha_d, \Xi_d$ and $\alpha_a, \Xi_a$ have a
more complicated form. They are given in appendix B, where we write also
the explicit expressions of the vector $w$ and the matrices ${\cal H}$
and ${\cal I}^c_K$
in the case of the $\lambda \Phi^4$ theory.
In spite of their form, the equations (\ref{3.3})-(\ref{3.6})
are coupled because the solutions $\alpha_b, \Xi_b$ and $\alpha_c, \Xi_c$
do not satisfy simple boundary conditions.
The expansion in powers of the sources of the
stationarity conditions (\ref{3.3}) and (\ref{3.6})
will provide approximate dynamical equations for the
expectations and correlations functions defined in eqs.
(\ref{2.3b})-(\ref{2.3f}).

\section{Expansion in powers of the sources}

\setcounter{equation}{0}

 We will use upper index $^{(0)}$ for the solutions of the dynamical equations
 when there are no sources.  The limit with vanishing source
corresponds to   $\alpha_a^{(0)}=0$ and $\Xi_a^{-1^{(0)}}=0$,
$
\alpha^{(0)}_b=\alpha^{(0)}_c=\alpha^{(0)}_d $
and
$
\Xi^{(0)}_b = \Xi^{(0)}_c = \Xi^{(0)}_d $;
we have also $w^b{}^{(0)}=w^c{}^{(0)}=w^{(0)}$ and
${\cal H}^b{}^{(0)}={\cal H}^c{}^{(0)} ={\cal H}^{(0)}$.
In this limit,  the dynamical equations (\ref{B.1}) and (\ref{B.2}) for
$\alpha_d$ et $\Xi_d$ become (we supress the index $d$) :
\be \label{5.3}
i \: \dot \alpha^{(0)} = \tau \: w^{(0)} \ , \ee
\be \label{5.4}
i \: \dot \Xi^{(0)} = - \left[ \left( \Xi^{(0)} + \tau \right)
\: {\cal H}^{(0)} \:
\left( \Xi^{(0)} - \tau \right) - \left( \Xi^{(0)} -
\tau \right) \: {\cal H}^{(0)} \:
\left( \Xi^{(0)} + \tau \right) \right] \ . \ee
These equations are the analog  for  the $\lambda \Phi^4$ theory of the
time-dependent Hartree-Bogoliubov (TDHB)
equations  in non-relativistic
physics. They are equivalent to the dynamical equations obtained in
reference \cite{5} where the authors used an alternative form of the
Balian-V\'en\'eroni variational
principle suited to the evaluation of single-time expectation values \cite{6}.

The first derivatives of
$W(J,t_0)=-i \: \ln {\cal Z}_{st}$
with respect to  the sources are equal to
 the expectations values with the index $c$. Indeed, the functional
 ${\cal Z}$ depend on the sources both explicitly and implicitly since
the approximate  solutions for ${\cal D}(t)$ and
${\cal A}(t)$ depend on the sources.
However  at the stationary point, only the explicit dependence
 contributes to the first derivative :
\be \label{6.2}
\frac{\delta {\cal Z}}{\delta J_j(t)} = i \: Tr \left( {\cal D}(t) {\cal A}(t)
Q_j \right) \ , \ee
which gives  for instance  :
$
\frac{\delta W }{\delta J_{\Phi}(\vec x,t)} = \varphi_c(\vec x,t)$.
The expressions for the second derivatives of $W$ are much more complicated.
The introduction of sources coupled to the composite operators
 $\Phi(\vec x) \Phi(\vec y), \Phi(\vec x) \Pi(\vec y)$ and
$\Pi(\vec x) \Pi(\vec y)$ together
with eq. (\ref{6.2}), allows us
to obtain dynamical equations for two-time correlation functions with three or
four field operators merely from  the expansion of the expectation values
$\alpha_c$ et $\Xi_c$ at the first order in powers of the sources.

{}From the first order corrections
$\alpha_c - \alpha^{(0)}$ and $\Xi_c-\Xi^{(0)}$,
we define the two-time correlation functions $\beta$ and $\Sigma$
($\beta$ is a vector
and $\Sigma$ is a matrix) :
\be \ba{ll} \label{6.7}
\alpha_c(\vec x,t)-
& {\displaystyle \alpha^{(0)}(\vec x,t) \simeq    i \: \int_{t_0}^{\infty}
dt'' \: \left \{
\int d^dy \: \beta^{\Phi}(\vec x, \vec y,t,t'') \: J^{\Phi}(\vec y,t'') +
  \beta^{\Pi} \:  J^{\Pi} \right.
} \\
&{\displaystyle  \left. +
 \int d^dx_1 d^dx_2 \: \beta^{\Phi \Phi}(\vec x, \vec x_1,\vec x_2,t,t'')
\: J^{\Phi\Phi}(\vec x_1,\vec x_2,t'')  + \beta^{\Phi \Pi} \: J^{\Phi \Pi}
+ \beta^{\Pi \Pi} \: J^{\Pi \Pi} \right \}
} \ea \ , \ee
\be \ba{ll} \label{6.8}
\Xi_c(\vec x, \vec y,t)-
& {\displaystyle \Xi^{(0)}(\vec x, \vec y,t) \simeq
 i \: \int_{t_0}^{+ \infty} dt''\: \left \{ \int d^dx_1 \:
\Sigma^{\Phi}(\vec x, \vec y,, \vec x_1,t, t'') \: J^{\Phi}(\vec x_1,t'') +
\Sigma^{\Pi} \: J^{\Pi} \right. } \\
& {\displaystyle \left. +   \int d^dx_1 d^dx_2 \:
\Sigma^{\Phi \Phi}(\vec x, \vec y,\vec x_1, \vec x_2,t, t'')
\: J^{\Phi \Phi}(\vec x_1,\vec x_2, t'') + \Sigma^{\Phi \Pi} \: J^{\Phi \Pi} +
\Sigma^{\Pi \Pi} \: J^{\Pi \Pi} \right \}
}  \ea \ . \ee
We have
\be \label{6.8a}
\beta_1^{\Phi \Phi}(\vec x,\vec y,\vec z,t,t") =
\Sigma^{\Phi}_{11}(\vec x, \vec y, \vec z, t,t") \ . \ee
The functions $\beta$ and $\Sigma$ provide
 approximations for the exact two-time
correlation functions $C^2$, $C^3$ and $C^4$ defined by eqs.
 (\ref{2.3d})-(\ref{2.3f}):
\be \label{6.8h}
 C^2_{\Phi \Phi}(\vec x, \vec y,t, t") \simeq
\beta^{\Phi}_1(\vec x, \vec y, t,t") \ , \ee
\be \ba{lll} \label{6.8i}
C^3(\vec x, \vec y, \vec z,t,t") \simeq
& {\displaystyle
\beta^{\Phi \Phi}_1(\vec x, \vec y, \vec z, t, t") } \\
& {\displaystyle -
 \varphi^{(0)}(\vec y,t") \:
\left(\beta_1^{\Phi}(\vec x, \vec z,t,t") + \varphi^{(0)}(\vec x,t)
\varphi^{(0)}(\vec z,t") \right)
} \\
& {\displaystyle -  \varphi^{(0)}(\vec z,t") \:
\left(\beta_1^{\Phi}(\vec x, \vec y,t,t") + \varphi^{(0)}(\vec x,t)
\varphi^{(0)}(\vec y,t") \right)  } \ea \ , \ee
\be \ba{lll} \label{6.8j}
C^4(\vec x, \vec y, \vec z,\vec u, t,t") \simeq
& {\displaystyle
\Sigma_{11}^{\Phi \Phi}(\vec x, \vec y, \vec z, \vec u,t,t") } \\
& {\displaystyle - \left( b_1^{\Phi}(\vec x, \vec z,t,t') +
\varphi^{(0)}(\vec x,t) \: \varphi^{(0)}(\vec z,t") \right) \:
 \left( b_1^{\Phi}(\vec y, \vec u,t,t') +
\varphi^{(0)}(\vec y,t) \: \varphi^{(0)}(\vec u,t") \right) } \\
& {\displaystyle - \left( b_1^{\Phi}(\vec x, \vec u,t,t') +
\varphi^{(0)}(\vec x,t) \: \varphi^{(0)}(\vec u,t") \right) \:
 \left( b_1^{\Phi}(\vec y, \vec z,t,t') +
\varphi^{(0)}(\vec y,t) \: \varphi^{(0)}(\vec z,t") \right) } \ea \ . \ee


The expansion in powers of the sources of the dynamical equations
(\ref{3.4}) and  (\ref{3.6}) for $\beta_c$ et $\Xi_c$ yields :
\be \label{6.9}
i \: \delta \dot \beta_c = \tau \: \left(\delta w^c- \delta w^c_K \right)
\ , \ee
\be \ba{ll} \label{6.10}
i \: \delta \dot \Xi_c = 2 \: & {\displaystyle  \left[
\delta \Xi_c \: {\cal H}^{(0)} \: \tau - \tau \: {\cal H}^{(0)} \:
\delta \Xi_c \right. } \\
& {\displaystyle \left.  + \Xi^{(0)} \: \left( \delta {\cal H}^c-
\delta {\cal I}^c_K
\right) \: \tau -
\tau \: \left( \delta {\cal H}^c- \delta {\cal I}^c_K \right) \:
\Xi^{(0)} \right] } \ea
\ . \ee
In these equations the matrix ${\cal H}$ has to be evaluated for the
TDHB solutions $\alpha^{(0)}, \Xi^{(0)}$ of eqs. (\ref{5.3})-(\ref{5.4}).
  The variations of
$w^c_K(\vec x,t)$ (eqs. (\ref{B.14})-(\ref{B.15}))
and  ${\cal I}^c_K(\vec x,\vec y,t)$ (eqs. (\ref{B.16})-(\ref{B.18}))
with respect to $J(t")$ will give  terms proportional to   $\delta(t-t")$.
{}From the variations   $\delta w^c$ and $\delta {\cal H}^c$,
we define the matrices
 $t_{ij}, T_{i,jk}, r_{ij,k},R_{ij,kl}$ (which are the analogs of
the  RPA kernel of  Balian and V\'en\'eroni \cite{1}) :
\be \label{6.11}
\delta w^c_i(\vec x,t)= t_{ij}(\vec x,t) \: \delta \alpha^c_j(\vec x,t)
-\frac{1}{2} \: \int d^dy \: T_{i,jk}(\vec x, \vec y,t) \: \delta
\Xi^c_{kj}(\vec y,\vec x,t) \ , \ee
\be \label{6.12}
\delta {\cal H}^c_{ij}(\vec x,\vec y,t)=r_{ij,k}(\vec x,\vec y,t) \:
\delta \alpha^c_k(\vec y,t) - \frac{1}{2} \: \int d^dz \:
R_{ij,kl}(\vec x, \vec z,t) \: \delta \Xi^c_{lk}(\vec z, \vec y,t)  \ . \ee
These matrices depend on the TDHB solutions  $\alpha^{(0)}$ and
$\Xi^{(0)}$.
Their expressions in the case of the $\lambda \Phi^4$ theory are given
in appendix B. With these notations, the dynamical equations for the two-time
and
three-point correlation functions $\beta^{\Phi}$ and $\Sigma^{\Phi}$ write :
\be \ba{ll} \label{6.14}
i \: \frac{d}{dt} \beta^{\Phi} (\vec x,\vec y,t,t")= \tau & {\displaystyle
\left[ t(\vec x,t) \: \beta^{\Phi}(\vec x,\vec y,t,t") -
\frac{1}{2} \: T_{,(jk}(\vec x, \vec x,t) \: \Sigma^{\Phi}_{kj)}
(\vec x, \vec x,
\vec y,t,t") \right. } \\
& {\displaystyle \left.
+ i \: \pmatrix{1 \cr 0 \cr } \delta(\vec x - \vec y) \:
\delta(t-t") \right] } \ea \ , \ee
\be \ba{ll} \label{6.14a}
i \: \frac{d}{dt} \Sigma^{\Phi}_{ij} (\vec x,\vec y,\vec x_1,t,t")=
& {\displaystyle 2 \: \left[ \Sigma^{\Phi} \: {\cal H}^{(0)} \: \tau
 - \tau \: {\cal H}^{(0)} \: \Sigma^{\Phi}  \right. } \\
& {\displaystyle \left.   + \Xi^{(0)} \: r_{,(k} \: \beta^{\Phi}_{k)} \: \tau
 - \tau \: r_{, (k} \: \beta^{\Phi}_{k)} \: \Xi^{(0)}  \right. } \\
& {\displaystyle \left. -\frac{1}{2} \: \Xi^{(0)}
 \: R_{,(kl} \: \Sigma^{\Phi}_{lk)} \:
\tau + \frac{1}{2} \: \tau \: R_{,(kl} \:
\Sigma^{\Phi}_{lk)} \Xi^{(0)} \right]_{ij}
} \ea \ . \ee
We obtain similar equations for the three-point functions
$\beta^{\Phi \Phi}, \beta^{\Phi \Pi}$,  $\beta^{\Pi \Pi}$
and the four-point functions
$\Sigma^{\Phi \Phi}, \Sigma^{\Phi \Pi}$ et $\Sigma^{\Pi \Pi}$.
These dynamical equations are not sufficient  since the boundary conditions
are
 $\alpha_a(t=+\infty)=0, \Xi_a^{-1}(t=+\infty)=0$ and
$\alpha_d(\vec x,t_0)=\alpha_0(\vec x), \:
 \Xi_d(\vec x, \vec y,t_0)=\Xi_0(\vec x,\vec y)$.
We will use the dynamical equations for the functions $l$ defined from the
expansion of $\Xi_a^{-1}$ at the first order :
\be \ba{llll} \label{6.16}
\Xi_a^{-1}(\vec x, \vec y,t) \simeq  &
{\displaystyle  i \: \int_{t_0}^{\infty}
dt'' \: \left \{
\int d^dx_1
\: l^{\Phi}(\vec x, \vec y,\vec x_1,t",t) \: J^{\Phi}(\vec x_1,t'')
 + l^{\Pi} \: J^{\Pi} \right.
} \\
&{\displaystyle  \left. +
 \int d^dx_1 d^dx_2 \: l^{\Phi \Phi}(\vec x, \vec y, \vec x_1,\vec x_2,t",t)
\: J^{\Phi\Phi}(\vec x_1,\vec x_2,t'')  + l^{\Phi \Pi} \: J^{\Phi \Pi}
+ l^{\Pi \Pi} \: J^{\Pi \Pi} \right \}
} \ea \ . \ee
We have also at the first order :
\be \label{6.17}
\alpha_c -\alpha_d \simeq \left( \tau \: \Xi_a^{-1} - \Xi_a^{-1} \: \Xi^{(0)}
\right) \: \alpha^{(0)}  \ee
and
\be \label{6.22}
\Xi_c- \Xi_d \simeq -\left(\Xi_d-\tau \right) \: \left( \Xi
^{(0)} \: \Xi_a^{-1} +
\Xi_a^{-1} \: \tau \right) \ . \ee
At the initial time $t_0$,
\be \label{6.20}
\alpha_c(t_0) -\alpha_0 \simeq \left( \tau \: \Xi_a^{-1} - \Xi_a^{-1} \: \Xi_0
\right) \: \alpha_0 \ , \ee
and
\be  \label{6.27}
\Xi_c(t_0)-\Xi_0 \simeq
 -\left(\Xi_0-\tau \right) \: \left( \Xi_0 \: \Xi_a^{-1}
+ \Xi_a^{-1} \: \tau \right)
 \ . \ee
This yields the following relations between the functions
$\beta^{\Phi}(t_0,t")$,
$\Sigma^{\Phi}(t_0,t")$ and the function $l^{\Phi}(t",t_0)$ :
\be \label{6.21}
\beta^{\Phi}(\vec x, \vec x_1, t_0,t")=
\int d^dz_1 \: d^dz_2 \:
\left( \tau \: l^{\Phi}(\vec x, \vec z_1, \vec x_1, t",t_0) \:
\delta(\vec z_1-\vec z_2) - l^{\Phi}(\vec x, \vec z_1, \vec x_1,t",t_0) \:
\Xi_0(\vec z_1, \vec z_2) \right) \: \alpha_0(\vec z_2) \ , \ee
\be \ba{ll}   \label{6.28}
\Sigma^{\Phi}(\vec x,\vec y,\vec x_1,t_0,t")=& {\displaystyle
 - \int d^dz_1 \: d^dz_2  \:
 \left( \Xi_0(\vec x, \vec z_1)-\tau \right)   } \\
& {\displaystyle \times
\left( \Xi_0(\vec z_1, \vec z_2) \: l^{\Phi}(\vec z_2,\vec y,\vec x_1, t",t_0)
+ \delta(\vec z_2-\vec y) \: l^{\Phi}(\vec z_1,\vec y,\vec x_1,t",t_0) \: \tau
\right) }  \ea \ . \ee
We have similar relations between the functions $\beta^{\Phi \Phi},
\Sigma^{\Phi \Phi}$ and $l^{\Phi \Phi}$.

To lowest order, the dynamical equation for $\Xi_a^{-1}$ reads :
\be  \label{6.29}
i \: \frac{d}{dt} \left( \Xi_a^{-1} \right) =
{\cal H}^c-{\cal H}^b-{\cal I}_K^c - 2 \: \Xi_a^{-1} \tau {\cal H}^{(0)} +
2 \: {\cal H}^{(0)} \tau \Xi_a^{-1}   \ .  \ee
{}From
the expansion  (\ref{6.16}) for $\Xi_a^{-1}$ and the variations
$\delta {\cal H}^c - \delta {\cal H}^b$, we obtain :
\be \ba{ll} \label{6.32}
i \: \frac{d}{dt} l(\vec x,\vec y,\vec x_1,t",t)_{ij}=
& {\displaystyle -2 \: r_{ij,k} \: \tau_{kl} \: l_{lm} \: \alpha_{m}^{(0)}
- R_{ij,kl} \: \left( \Xi^{(0)} \: l \: \tau - \tau \: \Xi^{(0)} \: l
\right)_{lk} } \\
& {\displaystyle + \left[ -2 \: l \tau {\cal H}^{(0)} + 2 \: {\cal H}^{(0)} \:
\tau \: l + i \: \delta {\cal I}^c_K \: \delta(t-t") \right]_{ij} } \ea \ . \ee
We have  $l(\vec x,\vec y,t",t)=0$ for  $t>t"$ and :
\be \label{6.33}
l^{\Phi \Phi}(\vec x, \vec y, \vec x_1,\vec x_2,t",t")=1 \: \: , \: \:
l^{\Phi \Pi}(\vec x, \vec y, \vec x_1,\vec x_2,t",t")=2i \: \: , \: \:
l^{\Pi \Pi}(\vec x, \vec y, \vec x_1,\vec x_2,t",t")=-1 \ , \ee
\be \label{6.34}
l^{\Phi}(\vec x, \vec y, \vec x_1,t",t")=
l^{\Pi}(\vec x, \vec y, \vec x_1,t",t")=0 \ . \ee

Therefore, to obtain approximations for the two-time causal
correlation functions
$C^2,C^3$ and $C^4$, we need first to solve
the TDHB equations (\ref{5.3}) and (\ref{5.4})
for $\alpha^{(0)}$ and $\Xi^{(0)}$ and to  evaluate the matrices
$r,R,t$ and $T$. Then we solve the equation (\ref{6.32}) for
$l^{\Phi}(\vec x, \vec y, \vec x_1,t",t)$
and
$l^{\Phi \Phi}(\vec x, \vec y, \vec x_1, \vec x_2,t",t)$ ($t">t'\ge t_0$)
backward from $t"$ to $t_0$ with the
boundary conditions (\ref{6.33})-(\ref{6.34}).
{}From the relations (\ref{6.21}) and (\ref{6.28}), we
obtain $\beta^{\Phi}(\vec x,\vec x_1,t_0,t"),
\beta^{\Phi \Phi}(\vec x,\vec x_1, \vec x_2,t_0,t")$ ,
$\Sigma^{\Phi}(\vec x, \vec y, \vec x_1,t_0,t")$ and
$\Sigma^{\Phi \Phi}(\vec x, \vec y, \vec x_1,\vec x_2,t_0,t")$ and
with these boundary conditions we solve
the dynamical equations for $\beta(t,t")$ and
$\Sigma(t,t")$ forward from $t$ to $t"$ .

In the TDHB approximation, the three-point function $C^3$ and the
four-point function $C^4$ vanish. Using the Balian-V\'en\'eroni
variational approach, we are able to obtain approximations for
$C^3$ and $C^4$ given by eqs. (\ref{6.8i}) and (\ref{6.8j}) which differ
from this na\"\i ve result.

In order to obtain approximations for the two-time anticausal functions,
it is sufficient to replace in the functional (\ref{2.4})-(\ref{2.5})
 the term $K_c$
by the analog term $K_b$. We can therefore derive a consistent
approximate formula for the response functions which involve the
retarded commutator.

When the two times co\"\i ncide in the function $\beta_1^{\Phi}$, we will
obtain an approximation for the function $G(\vec x, \vec x_1,t)$
which will differ from the TDHB solution $G^{(0)}$ (they satisfy however the
same initial condition
$G(\vec x, \vec x_1,t_0)=G^{(0)}(\vec x, \vec x_1,t_0)=G_0(\vec x, \vec x_1)$
).
We investigate
the static situation in a subsequent paper where we use a form of the
Balian-V\'en\'eroni variational principle which optimizes both the initial
state and the dynamics \cite{1}. We will  examine  the divergences
which appear in the variational equations and study how the approximations
obtained for the functions $G, C^3$ and $C^4$ are related to a few
physical quantities of the theory, i. e. the renormalized mass of the
particle and the renormalized coupling constant.

In conclusion, using an appropriate parametrization, we have shown how to
evaluate variationally
the generating functional for multi-time correlation functions
in a quantum field theory out of equilibrium
when the initial state is given and equal to a
Gaussian state. We have obtained  dynamical equations in
a compact form whose expansion at the first order in power of the sources
provides approximations for the two-time correlation functions.

\vspace*{1cm}

{\bf Aknowledgements}

We are grateful to   Hubert Flocard and Marcel V\'en\'eroni for very
entlightning
discussions and suggestions.

\vspace*{1cm}

{\bf Appendix A}

\setcounter{equation}{0}

\renewcommand{\theequation}{A.\arabic{equation}}

In this appendix we give the expressions of the expectation values
$\alpha_b, \alpha_c, \Xi_b$ and $\Xi_c$ as functions of
$\alpha_d, \alpha_a, \Xi_d$ and $\Xi_a$.

\be \label{A.1}
\alpha_b=\left( \Xi_a-\tau \right) \: \frac{1}{\Xi_a+\Xi_d} \:
\alpha_d + \left(\Xi_d + \tau \right) \: \frac{1}{\Xi_a+\Xi_d} \:
\alpha_a \ , \ee
\be \label{A.2}
\alpha_c = \left( \Xi_d - \tau \right) \: \frac{1}{\Xi_a + \Xi_d} \:
\alpha_a + \left( \Xi_a + \tau \right) \: \frac{1}{\Xi_a + \Xi_d} \:
\alpha_d \ , \ee
where  $\tau$ is the  $2 \times 2$ matrix :
\be \label{A.3}
\tau= \pmatrix{ 0 & 1 \cr -1 & 0 \cr } \ . \ee

For the matrices $\Xi$, we have the following relations :
\be \label{A.6}
\Xi_b- \tau = \left( \Xi_a -\tau \right) \: \frac{1}{\Xi_a +
\Xi_d} \: \left( \Xi_d - \tau \right) \ , \ee
\be \label{A.7}
\Xi_c - \tau = \left( \Xi_d - \tau \right) \:
\frac{1}{\Xi_a + \Xi_d} \: \left( \Xi_a - \tau \right) \ . \ee

\vspace*{1cm}

{\bf Appendix B}

\setcounter{equation}{0}

\renewcommand{\theequation}{B.\arabic{equation}}

In this appendix, we give the evolution equations for $\alpha_d, \Xi_d$ and
$\alpha_a, \Xi_a$.
\be \ba{ll} \label{B.1}
2 i \: \dot \alpha_d =
& {\displaystyle  - \left[ \left( \Xi_d + \tau \right) \left( {\cal H}^c
-{\cal I}^c_K \right) \left( \Xi_d - \tau \right) - \left( \Xi_d -
\tau \right) \: {\cal H}^b \: \left( \Xi_d + \tau \right) \right] \: \tau \:
\alpha_{b-c} } \\
& {\displaystyle + \Xi_d \left( w^c-w^b \right) + \tau \: \left( w^c+w^b
\right) - \left( \Xi_d + \tau \right) \: w^c_K } \ea  \ee
\be \label{B.2}
i \: \dot \Xi_d = - \left[ \left( \Xi_d+ \tau \right) \left( {\cal H}^c
-{\cal I}^c_K \right) \left( \Xi_d - \tau \right) - \left( \Xi_d -
\tau \right) \: {\cal H}^b \: \left( \Xi_d + \tau \right) \right]
\ee
\vspace*{0.5cm}
\be \ba{ll} \label{B.3}
2 i \: \dot \alpha_a =
& {\displaystyle   \left[ \left( \Xi_a - \tau \right) \left( {\cal H}^c
-{\cal I}^c_K \right) \left( \Xi_a + \tau \right) - \left( \Xi_a+
\tau \right) \: {\cal H}^b \: \left( \Xi_a - \tau \right) \right] \: \tau \:
\alpha_{c-b} } \\
& {\displaystyle + \Xi_a \left( w^b-w^c \right) + \tau \: \left( w^c+w^b
\right) - \left( \Xi_a - \tau \right) \: w^c_K } \ea \ , \ee
\be \label{B.4}
i \: \dot \Xi_a = - \left[ \left( \Xi_a- \tau \right) \left( {\cal H}^c
-{\cal I}^c_K \right) \left( \Xi_a + \tau \right) - \left( \Xi_a +
\tau \right) \: {\cal H}^b \: \left( \Xi_a - \tau \right) \right]
\ee
\vspace*{0.5cm}

The expressions for the vector $w$ and the matrices ${\cal H}$ and ${\cal I}$
are the following :
\be \label{B.5}
\tilde w^b(\vec x,t)_1 = \frac{\delta <H>_b}{\delta \varphi_b(\vec x,t)}
= - f^b(\vec x,t) \: \: \ , \: \:
\tilde w^b(\vec x,t)_2 = i \: \frac{\delta <H>_b}{\delta \pi_b(\vec x,t)}
= i \pi^b(\vec x,t) \ , \ee
\be \label{B.6}
{\cal H}^b(\vec x, \vec y,t)_{ij} =
-2 \frac{\delta <H>_b}{\delta \Xi_b(\vec x, \vec y,t)_{ji}} \ , \ee
\be \label{B.7}
{\cal H}^b(\vec x, \vec y,t)_{11}= - \frac{\delta <H>_b}{\delta
G_b(\vec y, \vec x,t)} = \frac{1}{2} \: g^b(\vec x,\vec y,t) \ , \ee
\be \label{B.8}
 {\cal H}^b(\vec x, \vec y,t)_{22}= + \frac{\delta <H>_b}{\delta
S_b(\vec y, \vec x,t)} = \frac{1}{2} \: \delta(\vec x-\vec y) \ , \ee
\be \label{B.9}
 {\cal H}^b(\vec x, \vec y,t)_{12}= 2 i \: \frac{\delta <H>_b}{\delta
T_b(\vec y, \vec x,t)} = 0\ . \ee
For a self-interacting scalar field, we have :
\be \label{B.10}
f_b(\vec x,t) =- \left( -\Delta + m_0^2 +\frac{b}{6} \: \varphi^2_b(\vec x,t)
+ \frac{b}{2} \: G_b(\vec x, \vec x,t)\right) \: \varphi_b(\vec x, t) \ , \ee
\be \label{B.11}
g_b(\vec x,\vec y,t) = - \left( - \Delta + m_0^2 + \frac{b}{2} \:
\varphi^2_b(\vec x,t) + \frac{b}{2} \: G_b(\vec x,\vec x,t) \right) \:
\delta(\vec x- \vec y) \ . \ee
{}From the variations of the source term $K_c$, we obtain :
\be \label{B.12}
\tilde w^c_K(\vec x,t)_1= \frac{\delta K_c}{\delta \varphi_c(\vec x,t)} \: \:
\ , \: \:
\tilde w^c_K(\vec x,t)_2= i \: \frac{\delta K_c}{\delta \pi_c(\vec x,t)} \ ,
\ee
\be \label{B.13}
{\cal I}^c_K(\vec x, \vec y,t)_{ij} = -2 \frac{\delta K_c}{\delta
\Xi_c (\vec y, \vec x,t)_{ji}} \ . \ee
\be \label{B.14}
w^c_K(\vec x,t)_1=J^{\Phi}(\vec x,t) + 2 \: \int d^dx_2 \:  \left(
J^{\Phi \Phi} (\vec x, \vec x_2,t) \: \varphi_c(\vec x_2,t) +
J^{\Phi \Pi} (\vec x, \vec x_2,t) \: \pi_c(\vec x_2,t) \right) \ , \ee
\be \label{B.15}
w^c_K(\vec x,t)_2=i \: J^{\Pi}(\vec x,t) + 2  \: i \: \int d^dx_2  \: \left(
J^{\Phi \Pi} (\vec x, \vec x_2,t) \: \varphi_c(\vec x_2,t) +
J^{\Pi \Pi} (\vec x, \vec x_2,t) \: \pi_c(\vec x_2,t) \right) \ , \ee
\be \label{B.16}
{\cal I}^c_K(\vec x, \vec y,t)_{11}= -J^{\Phi \Phi} (\vec x, \vec y,t) \ , \ee
\be \label{B.17}
{\cal I}^c_K(\vec x, \vec y,t)_{12}= -2 i \:
J^{\Phi \Pi} (\vec x, \vec y,t) \ , \ee
\be \label{B.18}
{\cal I}^c_K(\vec x, \vec y,t)_{22}= J^{\Pi \Pi} (\vec x, \vec y,t) \ . \ee

The expressions for the matrices $t, T,r$ and $R$ which appear in the
dynamical equations for the two-time correlation functions are the following :
\be \label{B.19}
t_{11}(\vec x, t)=
-g^{(0)}(\vec x,\vec x,t) \: \: , \: \: t_{22}=-1 \ , \ee
\be \label{B.20}
T_{1,11}(\vec x, \vec y,t) =
\frac{b}{2} \: \varphi^{(0)}(\vec x,t) \delta(\vec x-\vec y)
\: \: , \: \: T_{2,jk}=0 \ , \ee
\be \label{B.21}
r_{11,1}(\vec x,\vec y,t) = - \frac{b}{2} \: \varphi^{(0)}(\vec x,t)
\delta(\vec x-\vec y) \ , \ee
\be \label{B.22}
R_{11,11}(\vec x,\vec y,t) = \frac{b}{4} \: \delta(\vec x-\vec y) \ . \ee
The other matrix elements are equal to zero.

\end{document}